\documentclass[a4paper,12pt]{article}

\usepackage[utf8]{inputenc}
\usepackage{cancel}
\usepackage{tikz}
\usepackage{ulem}
\usepackage{amsfonts}
\usepackage{amssymb}
\usepackage{graphicx}
\usepackage{amsmath}
\usepackage{enumerate}
\usepackage{mathtools}
\usepackage{subfig}
\usepackage{color}
\usepackage{tikz}
\usepackage{float}
\usepackage{here}
\usepackage{cite}
\usepackage{mathrsfs}
\usepackage{float,epsfig}
\usepackage{dcolumn}
\usepackage{graphicx}
\usepackage{bm}
\usepackage{amsmath,amssymb,amsthm}
\usepackage[colorlinks=true,linkcolor=blue,citecolor=red]{hyperref}
\usepackage{multirow}
\usepackage[toc,page]{appendix}
\usepackage{booktabs}   
\usepackage{multirow}   
\usepackage{array}
\usetikzlibrary{arrows.meta}
\usetikzlibrary{bending}
\usetikzlibrary{calc}
\newcommand{\be}{\begin{equation}}
\newcommand{\ee}{\end{equation}}
\newcommand{\bea}{\setlength\arraycolsep{2pt} \begin{eqnarray}}
\newcommand{\eea}{\end{eqnarray}}

\def\0{{\sst{(0)}}}
\def\1{{\sst{(1)}}}
\def\2{{\sst{(2)}}}
\def\3{{\sst{(3)}}}
\def\4{{\sst{(4)}}}
\def\5{{\sst{(5)}}}
\def\6{{\sst{(6)}}}
\def\7{{\sst{(7)}}}
\def\8{{\sst{(8)}}}
\def\sst#1{{\scriptscriptstyle #1}}

\makeatletter \@addtoreset{equation}{section}

\definecolor{lime}{HTML}{A6CE39}

\begin{document}
\title{{\normalsize \textbf{\Large Constrained Scenarios of Non-Commutative Schwarzschild Black Holes with Global Monopoles from Van der Waals Behaviors    }}}
\author{ {\small  Ismail Benyaich$^1$  and Maryem  Jemri$^2$\thanks{Corresponding author: maryem.jemri@um5r.ac.ma} \footnote{
Authors are listed  in alphabetical order. 
} \hspace*{-8pt}} \\
{\small
$^1$ Faculty of Science, Mohammed V University in Rabat, Rabat, Morocco } \\
{\small $^2$ ESMaR, Faculty of Science, Mohammed V University in Rabat, Rabat, Morocco   }}
\maketitle

\begin{abstract}
Combining thermodynamics and  optics, we investigate the  Van  der Waals behaviors of non-commutative Schwarzschild black holes with global monopoles. First, we examine the constrained thermodynamic recovering such properties. Using these thermodynamic results,  we approach the shadows  by merging  Event Horizon Telescope findings and the Van der Waals limits. 
 In this context, we provide a machine learning study to  determine the compatibility with the empirical data  of $M87^*$, $SgrA^*_{\mathrm{VLTI}}$, and $SgrA^*_{\mathrm{Keck}}$ black holes. As a result, we find that  such a proposed model matches with  the $M87^*$ observational data.

\end{abstract}

\newpage

\section{Introduction}
Recently, the black holes in the Anti-de Sitter (AdS) spacetime geometries have received a remarkable interest. In particular, the thermodynamics of such solutions with a negative cosmological constant have been extensively investigated using different methods \cite{i1}. Identifying  the  cosmological constant with the pressure thermodynamic quantity, various phase transitions and criticality features of AdS black holes   have been studied  showing interesting results  \cite{i11}. A special focus has been on the Reissner-Nordström-AdS (RN-AdS) solutions, where the $(P-v)$ criticality exhibits relevant  properties.  Approaching the state equation $P=P(T,v)$, it has been shown that such solutions develop similarities with the Van der Waals systems\cite{i3}. Precisely, the quantity $
\chi=\frac{P_c v_c}{T_c}$
representing the critical ratio has been found to be
$ \frac{3}{8}$
evaluated at the critical point of the phase transitions\cite{i4,i44}. This universal value has shown a nice  interplay between charged AdS black holes  and fluid systems. However, this number has been modified by considering external parameters derived via different scenarios, including modified gravity theories\cite{i5,i6}.

More recently, a non-commutative (NC) Schwarzschild solution with a term showing similarities with ordinary charged solutions have been reconsidered\cite{i7}. In particular, the term $
\frac{8\pi M \Theta}{\sqrt{\pi}\,r^2}
$
has been added to the radial function of the ordinary Schwarzschild solutions\cite{i8}. The associated criticality behaviors have been examined, providing a universal value
$
\chi = 0.36671
$
being different to the usual one  identified with  $3/8$ associated with the Van der Waals behaviors\cite{i9}. A close examination has shown that such a universal number could be recovered by implementing certain extra external parameters.

The aim of this work is to contribute to  these activities by reconsidering the study of such deformed Schwarzschild solutions. The objective is twofolds. First, we recover the Van der Waals-like behaviors in NC   Schwarzschild black holes by adding global  monopole  (GM) contributions. 
Second, we approach the optical  properties  for such universal critical behaviors. More precisely, we constrain  the associated parameters by combining machine learning techniques and empirical data of the Event Horizon Telescope (EHT) findings to  classify  whether the resulting parameter space is consistent with the observations of $M87^*$, $SgrA^*_{\mathrm{VLTI}}$, and $SgrA^*_{\mathrm{Keck}}$ black holes. As a result,  we find that the proposed model matches with the $M87^*$ empirical data 

The organization of the paper is as follows. In Section 2, we briefly discuss the NC Schwarzschild black hole with GMs.  In Section 3, we investigate the thermodynamic  critical behavior and the Joule--Thomson effect in the van der Waals-like limits. In Section 4,  we examine the black hole shadow and compare the theoretical predictions with observational data from the EHT international  collaborations for such thermodynamic  critical limits.  Section 5 exposes  the machine learning methods matching  with the $M87^*$, $SgrA^*_{\mathrm{VLTI}}$, and $SgrA^*_{\mathrm{Keck}}$ observational data. Finally, the last section is devoted to concluding remarks.

\section{NC Schwarzschild black hole with GMs}
In this section, we consider a static and spherically symmetric black hole in a NC geometry in the presence of  GMs \cite{b1,b2}.  It has been  remarked  that NC geometries  have been extensively  studied  in connections with D-brane physics and superstring models.  This can be modeled by open strings moving on D-branes in the presence of the antisymmetric  stringy field $B$ of  the NS-NS sector in type II superstrings\cite{b3,b4}.  Roughly,  it has  been shown that the NC effects are introduced through a smeared distribution of the black-hole mass, characterized by the coordinate commutation relation
\begin{equation}
[x^\mu,x^\nu]=
i\Theta^{\mu\nu},
\label{eq:nc_commutator}
\end{equation}
where $\Theta^{\mu\nu}$ is a constant antisymmetric tensor. For simplicity reasons, one can consider 
\begin{equation}
\Theta^{\mu\nu}=\Theta\epsilon^{\mu\nu},
\end{equation}
where $\epsilon^{\mu\nu}$ is the totally antisymmetric tensor and $\Theta$ is a  NC parameter with dimension $[\Theta]=[L^2]$ in such inspired string theory models. In the strong-field regime, this tensor has been shown to be closely related to the inverse of the antisymmetric  stringy field $B$ of the  NS-NS sector, which is regarded as a fundamental entity in string theory and related topics \cite{b3,b4}.  In this scenario, the corresponding static and spherically symmetric line element is given by
\begin{equation}
ds^2=
-f(r),dt^2
+\frac{dr^2}{f(r)}
+r^2\left(d\theta^2+\sin^2\theta d\phi^2\right),
\label{eq:nc_gm_metric}
\end{equation}
with the metric radial  function
\begin{equation}
f(r)=
1
-\frac{2\mathcal{M}_{\Theta}(r)}{r}
-\frac{\Lambda}{3}r^2-8\pi\xi\eta^2,
\label{eq:f_mass}
\end{equation}
where $\Lambda$ denotes the cosmological constant, while the GM parameters $\xi$ and $\eta$ represent the sign of the scalar-field kinetic term and the energy scale of symmetry breaking, respectively. For $\xi=1$, the scalar field has a positive kinetic energy, corresponding to an ordinary GM (OGM), whereas $\xi=-1$ leads to a negative kinetic energy and the formation of a phantom GM (PGM). For simplicity, we set $\xi=1$ throughout this work. The radial mass function $\mathcal{M}_{\Theta}(r)$ is determined by the smeared mass density according to
\begin{equation}
\mathcal{M}_{\Theta}(r)=
4\pi\int_{0}^{r}
\rho_{\Theta}(r')r'^2dr'.
\label{eq:mass_distribution}
\end{equation}
In this framework,   the function $\rho_{\Theta}(r)$ is modeled by a Lorentzian distribution of the form
\begin{equation}
\rho_{\Theta}(r)=
\frac{M\sqrt{\Theta}}
{\pi^{3/2}(r^2+\Theta)^2},
\label{eq:rho_theta}
\end{equation}
where $M$ represents the total black hole mass. In the large-$r$ regime, the corresponding mass function can be expanded as
\begin{equation}
\mathcal{M}_{\Theta}(r)=
M-\frac{4M\sqrt{\Theta}}
{\sqrt{\pi}r}
+\mathcal{O}\left(\Theta^{3/2}\right).
\label{eq:mass_expansion}
\end{equation}
Substituting Eq.~(\ref{eq:mass_expansion})  into Eq.~(\ref{eq:f_mass}),  one gets 
\begin{equation}
f(r)=
1
-\frac{2M}{r}
+\frac{aM}{r^2}
-\frac{\Lambda}{3}r^2-8\pi\eta^2,
\label{eq:f_metric}
\end{equation}
where one has used a new parameter 
\begin{equation}
a=\frac{8\sqrt{\Theta}}{\sqrt{\pi}},
\label{eq:a_theta}
\end{equation}
which quantifies the strength of the NC correction and its dimension is $[a]=[L]$.

In the commutative limit, $\Theta=0$, or equivalently $a=0$, the metric reduces to the AdS black hole with a GM. Conversely, setting $\eta=0$ eliminates the GM contribution and recovers the corresponding NC black-hole geometry\cite{i9}. These limits provide consistency checks for the solution and allow the effects of noncommutativity and the topological defect to be distinguished.

 Having discussed the proposed model,  we move now to examine   the  associated Van der Waals behaviors   by combining  the thermodynamic and the optical properties in the forthcoming sections.

\section{Thermodynamic Van der Waals behaviors }
In this section, we investigate the   thermodynamic of the proposed black hole models  at  Van der Waals limits. Indeed, the thermodynamic properties of the black hole can be obtained directly from the metric function $f(r)$. In particular, the event-horizon condition $f(r_h)=0$ allows the black hole mass to be expressed in terms of the horizon radius as
\begin{equation}
M =
\frac{r_h^{2} \left(24 \pi \eta^{2}+\Lambda r_h^{2}-3\right)}
{3(a-2r_h)}.
\end{equation}
In the absence of the NC and GM  parameters, this expression reduces to the Schwarzschild--AdS result identified with 
$
M=\frac{3r_h^{2}-\Lambda r_h^{4}}{6r_h}
$. The Hawking temperature is obtained from the surface gravity $
T_H=\frac{\kappa}{2\pi},$
where the surface gravity is given by
$
\kappa=
\left.
\frac{1}{2}\frac{\partial f(r)}{\partial r}
\right|_{r=r_h},
$
which yields
\begin{equation}
T_H=
\frac{
\left(r_h-a\right)
\left(24\pi\eta^{2}+\Lambda r_h^{2}-3\right)r_h^{2}
-\Lambda r_h^{4}(a-2r_h)
}
{6\pi(a-2r_h)r_h^{3}}.
\label{eq:TH}
\end{equation}
To investigate the thermodynamic stability, we further calculate two important quantities, namely the Gibbs free energy and the heat capacity, using the entropy $S$. The latter is obtained from the Bekenstein--Hawking area law,
$
S=\frac{\mathcal{A}}{4}=\pi r_h^2,
$
where
$
\mathcal{A}
=
\int\int
\sqrt{g_{\theta\theta}g_{\phi\phi}}
\,d\theta\,d\phi
=
4\pi r_h^2
$
is the area of the black hole event horizon. In this way,  the Gibbs free energy is given by
\begin{equation}
G=
\frac{\Lambda r_h^3(2a-r_h)
+3\left(8\pi\eta^{2}-1\right)r_h^{2}
+3\left(8\pi\eta^{2}-1\right)ar_h
}
{6(a-2r_h)}.
\end{equation}
It has been revealed that the sign of the Gibbs free energy characterizes the global thermodynamic stability. In particular, $G<0$ corresponds to a globally stable configuration, whereas $G>0$ indicates the global instability. The local thermodynamic stability can instead be studied  via  the heat capacity at constant pressure given by
\begin{equation}
C_P=
T_H\left(\frac{\partial S}{\partial T_H}\right)_P.
\end{equation}
Using the Hawking temperature, we obtain
\begin{equation}
C_P=-
\frac{
4 \pi r_h^2(3(a-r_h)(8\pi \eta^2-1)+\Lambda r_h^2(2a-3r_h))
}
{
3(8 \pi \eta^2-1)(2r_h^2-4ar_h+a^2)+2 \Lambda r_h^2(3(r_h^2-ar_h)+a^2)
}.
\end{equation}
The sign of the heat capacity determines the local thermodynamic stability of the black hole. A configuration with $C_P>0$ is locally  stable, whereas $C_P<0$ indicates an unstable configuration. These thermodynamic behaviors can be directly illustrated through their graphical representations.    Since they are not relevant in   the present discussion,  such graphs are omitted. 

We next investigate the thermodynamic universalities of the charged NC black hole with  GMs that we are after. Within the extended phase-space formalism, the cosmological constant is interpreted as a thermodynamic pressure expressed as 
\begin{equation}
P=-\frac{\Lambda}{8\pi}.
\end{equation}
Using Eq.~\eqref{eq:TH}, the corresponding equation of state is obtained as
\begin{equation}
P=
\frac{
3(8\pi \eta^{2}-1)(a-r_h)+6\pi Tr_h(a-2r_h)
}
{
8\pi r_h^{2}(2a-3r_h)
}.
\label{eq:P}
\end{equation}
This equation of state provides the basis for investigating the thermodynamic universalities through  the $P-V$ criticality and the Joule--Thomson effect. The critical point is determined by the conditions
\begin{equation}
\frac{\partial P}{\partial v}=0,
\qquad
\frac{\partial^{2}P}{\partial v^{2}}=0,
\end{equation}
where $v$ denotes the specific volume. The corresponding critical pressure,  the temperature, and  the specific volume are found to be 
\begin{align}
P_c&=
\frac{-0.0687069522\,\eta^{2}+0.002733762781}{a^{2}},
\\
T_c&=
\frac{-0.9149013775\,\eta^{2}+0.03640276923}{a},
\\
v_c&=
4.883109171\,a.
\end{align}
The useful dimensionless quantity characterizing the critical behavior is shown  to be 
\begin{equation}
\chi=
\frac{ 0.33550355 \eta^{2}- 0.013349262}{ 0.91490138 \eta^{2}- 0.036402769}.
\end{equation}
At this level, we provide certain comments on such a   critical ratio.
Taking $\eta=0$, this ratio becomes
\begin{equation}
\chi=0.3667100716,
\end{equation}
which is consistent with the result reported in  ~\cite{i9}. However, this value does not exactly reproduce the universal Van der Waals ratio being the numerical value  $3/8$ \cite{i2,ii2}. Therefore, we impose a constraint on the GM parameter $\eta$ such that the Van der Waals behavior is recovered. Solving  the algebraic equation  $\chi=\frac{3}{8}$ describing  the Van der Waals behaviors, we obtain
\begin{equation}
\eta^2=0.039788736.
\label{eq:constraint1}
\end{equation}
This value of the GM parameter exactly reproduces the Van der Waals behavior through the $P$-$v$ criticality, independently of the value of the NC parameter. To further investigate the significance of this specific GM parameter value in recovering the Van der Waals behavior, we consider the Joule--Thomson expansion, which describes the temperature variation of a system during an isenthalpic expansion\cite{i21}. It is characterized by the Joule--Thomson coefficient
\begin{equation}
\mu=
\left(\frac{\partial T}{\partial P}\right)_M
=
\frac{1}{C_P}
\left[
T\left(\frac{\partial V}{\partial T}\right)_P-V
\right].
\end{equation}
The inversion curve is determined by the condition $\mu(T_i)=0$. The corresponding inversion temperature is found to be 
\begin{equation}
T_i=
\frac{
16P_i \pi ar_i^{2}(a^2-3r_i(a-r_i))+3(a^2-2r_i(2a-r_i))(8\pi \eta^2-1)}
{
18\pi r_i(a-2r_i)^{2}
},
\label{eq:Ti}
\end{equation}
where $P_i$ denotes the inversion pressure. The equation of state can also be written in terms of the inversion pressure as
\begin{equation}
T=
\frac{8\pi P r^2(2a-3r)+3(r-a)(8\pi \eta^2-1)
}
{
6\pi r(a-2r)
}.
\end{equation}
Combining these expressions at the inversion point gives
\begin{equation}
3 a r_i \left(40 P_i \pi  \,r_i^{2}-13(8 \pi  \eta^{2}-1)\right)-4 a^{2} \left(8 P_i \pi  \,r_i^{2}-3(8 \pi  \eta^{2}-1)\right)-24 r_i^{2} \left(4 P_i \pi  \,r_i^{2}-8 \pi \eta^{2}+1\right)=0.
\end{equation}
 The minimum inversion temperature can be expressed as
\begin{equation}
T_i^{\min}
=
\frac{-64\pi\eta^{2}+8}
{\left(5+\sqrt{41}\right)^{2}\pi a}.
\end{equation}
The ratio between the minimum inversion temperature and the critical temperature is defined as
\begin{equation}
\zeta=\frac{T_i^{\min}}{T_c}.
\end{equation}
For the present black-hole system, this ratio becomes
\begin{equation}
\zeta=
\frac{ 0.4921894061 \eta^{2}- 0.01958359423}{ 0.9149013775 \eta^{2}- 0.03640276923}.
\end{equation}
For $\eta=0$, this ratio does not reproduce neither the corresponding RN--AdS result, $\zeta=1/2$, nor the Van der Waals value, $\zeta=3/4$\cite{i22}. Taking $\eta^2=0.039788736$,  however, as obtained independently from the Van der Waals behavior through the $P$-$V$ criticality analysis, we  find 
\begin{equation}
\zeta\simeq\frac{3}{4},
\end{equation}
which  recovers the Van der Waals behavior via the Joule–Thomson effect. Interestingly, this  numerical value of  the parameter $\eta^2$ agrees with the constraint obtained independently from the $P$-$V$ criticality analysis. Therefore, at $\eta^2=0.039788736$, the black-hole system reproduces the characteristic thermodynamic behavior of a Van der Waals fluid, as confirmed independently by both the $P$-$V$ criticality and the Joule--Thomson inversion analyses.

We could expect a connection between the thermodynamic and optical behaviors of the black hole at the Van der Waals limit, providing a link between its thermodynamic properties and the optical signatures accessible through the EHT observations. 
This motivates the investigation of the black hole optical behavior at  such a  critical value of $\eta^2$, which reproduces Van der Waals-like behaviors, with particular emphasis on the corresponding shadow properties.
\section{Constraining parameters from optical properties at Van der Waals limits}
In this section, we investigate the shadow of the black hole solutions at the thermodynamically distinguished value of the GM parameter which recovers the Van der Waals limit, and examine how the model parameters affect the radius of the photon sphere and the resulting shadows. To describe  the photon dynamics in the black hole spacetime, we employ the Hamilton–Jacobi formalism 
\begin{equation}
\frac{\partial S}{\partial \sigma}+H=0,
\end{equation}
where $S$ and $\sigma$ denote the Jacobi action and the affine parameter along the photon geodesics, respectively \cite{1,2}. For a massless photon propagating in a static and spherically symmetric spacetime, the Hamiltonian is given by
\begin{equation}
H=\frac{1}{2}g^{ij}p_i p_j=0.
\end{equation}
Due to the spherical symmetry of the spacetime, the photon motion can be restricted to the equatorial plane, $\theta=\pi/2$. The Hamiltonian equation then takes the form
\begin{equation}
\frac{1}{2}\left[-\frac{p_t^2}{f(r)}+f(r)p_r^2+\frac{p_\phi^2}{r^2}\right]=0.
\end{equation}
Since the Hamiltonian is independent of the coordinates $t$ and $\phi$, the corresponding conserved quantities are defined as
\begin{equation}
p_t=-E,\qquad p_\phi=L,
\end{equation}
where $E$ and $L$ represent the energy and the  angular momentum of the photon, respectively. The Hamiltonian equations of motion consequently yield
\begin{equation}
\dot t=-\frac{p_t}{f(r)},\qquad
\dot r=f(r)p_r,\qquad
\dot\phi=\frac{p_\phi}{r^2},
\end{equation}
where the dot denotes differentiation with respect to the affine parameter $\sigma$. Combining these relations with the null-geodesic condition gives the radial equation
\begin{equation}
\dot r^{2}+V_{\rm eff}(r)=0,
\end{equation}
with the effective potential
\begin{equation}
V_{\rm eff}(r)=f(r)\left[ \frac{L^2}{r^2}-\frac{E^2}{f(r)}\right].
\end{equation}
For a circular photon orbit, the radial motion must satisfy
\begin{equation}
V_{\rm eff}(r_p)=0,\qquad
\left.\frac{dV_{\rm eff}}{dr}\right|_{r=r_p}=0,
\end{equation}
where $r_p$ denotes the photon-sphere radius. These conditions lead to the following algebraic equation
\begin{equation}
6 \Lambda  \,r_p^{5}-4 a \Lambda  \,r_p^{4}+6\left(8 \pi  \,\eta^{2}-1\right) r_p^{3}-6\left(8 \pi  \,\eta^{2}-1\right) a \,r_p^{2}=0.
\end{equation}
Since this equation is generally difficult to solve analytically, the photon-sphere radius is determined numerically for the parameter values considered below. The relevant outer solution corresponds to the unstable circular photon orbit. Roughly, the photon trajectory can be obtained from
\begin{equation}
\frac{dr}{d\phi}
=\frac{\dot r}{\dot\phi}
=\frac{r^2 f(r)}{L}p_r.
\end{equation}
With the help of the conserved quantities,  it  can be  expressed  as
\begin{equation}
\frac{dr}{d\phi}
=\pm r\sqrt{f(r)\left(  \frac{r^2E^2}{f(r)L^2}-1  \right)}.
\end{equation}
At the turning point $\dfrac{dr}{d\phi}_{r=R}=0$, the impact parameter is fixed by the corresponding condition. This gives
\begin{equation}
\frac{dr}{d\phi}
=\pm r\sqrt{f(r)\left( \frac{r^2f(R)}{f(R)R^2}-1  \right)}.
\end{equation}
For a static observer located at $r_0$, the angle $\vartheta$ between the incoming photon and the radial direction satisfies 
\begin{equation}
\frac{\sqrt{g_{rr}}}{g_{\phi\phi}}
\left.\frac{dr}{d\phi}\right|_{r=r_0}
= \cot\vartheta.
\end{equation}
Consequently, the angular radius of the black hole shadow is obtained as

\begin{equation}
R_s=r_0\sin\vartheta
=R\sqrt{\frac{f(r_0)}{f(R)}}\bigg|_{R=r_p}.
\end{equation}
Thus, the shadow radius is directly determined by the radius of the unstable photon sphere and the metric function evaluated at the observer position. In the asymptotic observer limit, the apparent shadow can also be described through the celestial coordinates 
\begin{align}
x=&\lim_{r_0\rightarrow\infty}
\left(-r_0^2\sin\theta_0\frac{d\phi}{dr}\mid_{(r_0,\theta_0)}\right),\\
y=&\lim_{r_0\rightarrow\infty}
\left(r_0^2\frac{d\theta}{dr}\mid_{(r_0,\theta_0)}\right)
\end{align}
where $(r_0,\theta_0)$ are the position coordinates of the observer.  It is  denoted that the coordinate $X$ measures the apparent horizontal displacement of the image, whereas $Y$
corresponds to the vertical displacement in the observer sky.

\begin{figure}[h!]
\centering
\begin{tabular}{c}
\includegraphics[scale=0.4]{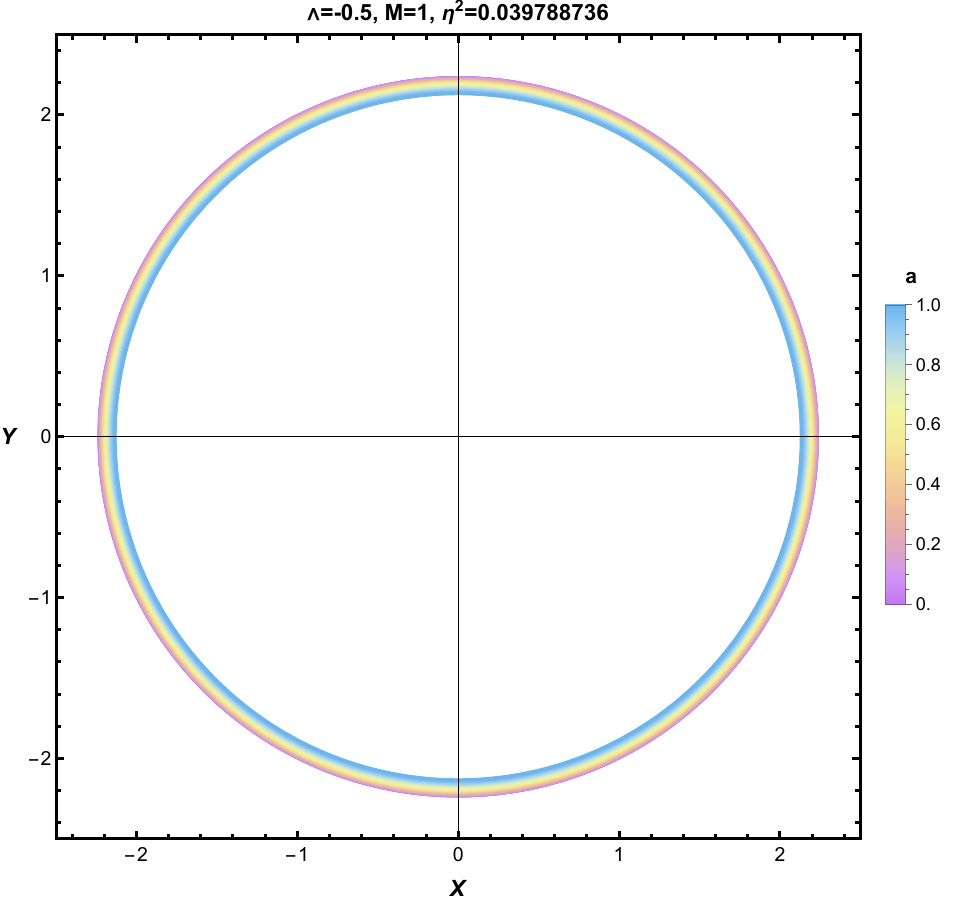}
\end{tabular}
\caption{Black hole shadows for different values of the NC parameter, verifying the Van der Waals limits.}
\label{sh1}
\end{figure}

We now examine the influence of the black hole parameters on the shadow radius. In particular, we investigate the effect of the NC parameter by fixing $\eta^2=0.039788736$, being  a value motivated by the thermodynamic analysis, which indicates that the proposed black holes exhibit the Van der Waals-like behaviors. To this end, we employ a numerical approach to compute the shadow contours over a broad range of the parameter values. All other parameters are kept fixed, while the NC parameter is varied with a step size of $0.01$. Fig. (\ref{sh1}) illustrates the dependence of the black hole shadow on the NC parameter for  such a  fixed value of the GM parameter recovering the    Van der Waals-like limits in  the thermodynamic description.

Fig. (\ref{sh1}) shows that increasing the NC parameter modifies the shadow radius while preserving its circular shape as expected since we are dealing with non rotating  solutions. To connect the theoretical predictions with observational data in the Van der Waals regime, we further confront the resulting shadow radius with the observations reported by the  EHT collaborations \cite{4}. In particular, we consider the observations of $M87^*$ and $SgrA^*$ and use them to constrain the parameters of the black hole model \cite{AE1,AE2}.  Following the observational analysis, we  consider the deviation of the shadow size from the Schwarzschild prediction through
\begin{equation}
d=\frac{R_s}{r_{\rm sh}}-1,
\end{equation}
where $R_s$ is the shadow radius and $r_{\rm sh}$  represents  the corresponding Schwarzschild shadow radius. The dimensionless ratio $R_s/M$ is then used as the main observable for confronting the theoretical model with the EHT measurements. The corresponding $1-\sigma$ and $2-\sigma$ observational intervals are summarized in Table~\ref{t1}.

\begin{figure}[h!]
\centering
\includegraphics[scale=0.3]{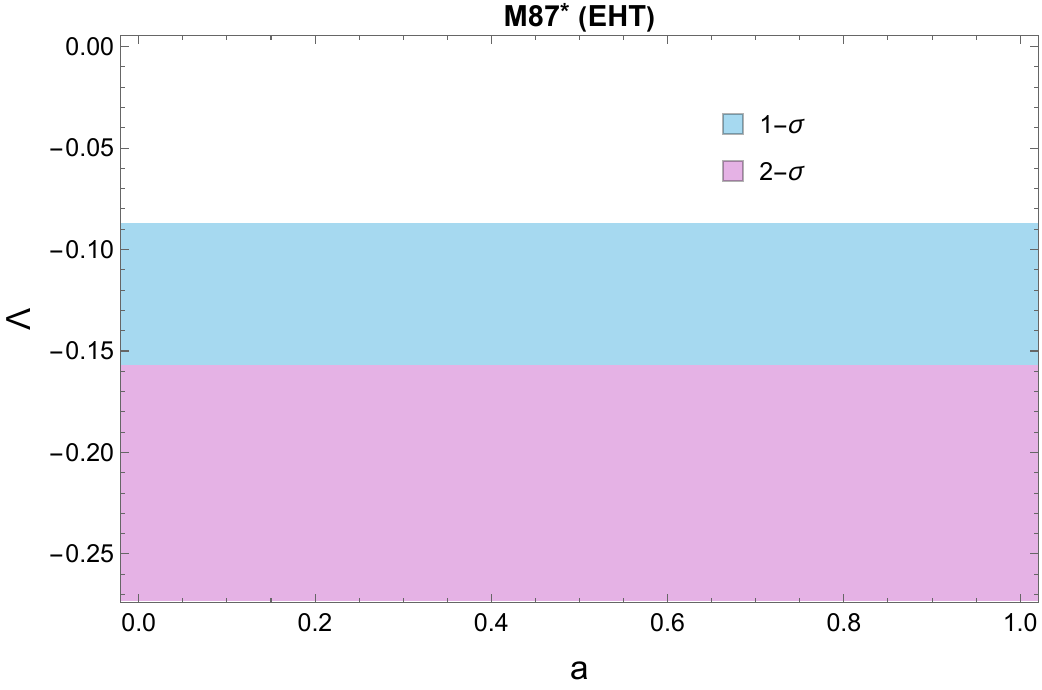} \includegraphics[scale=0.3]{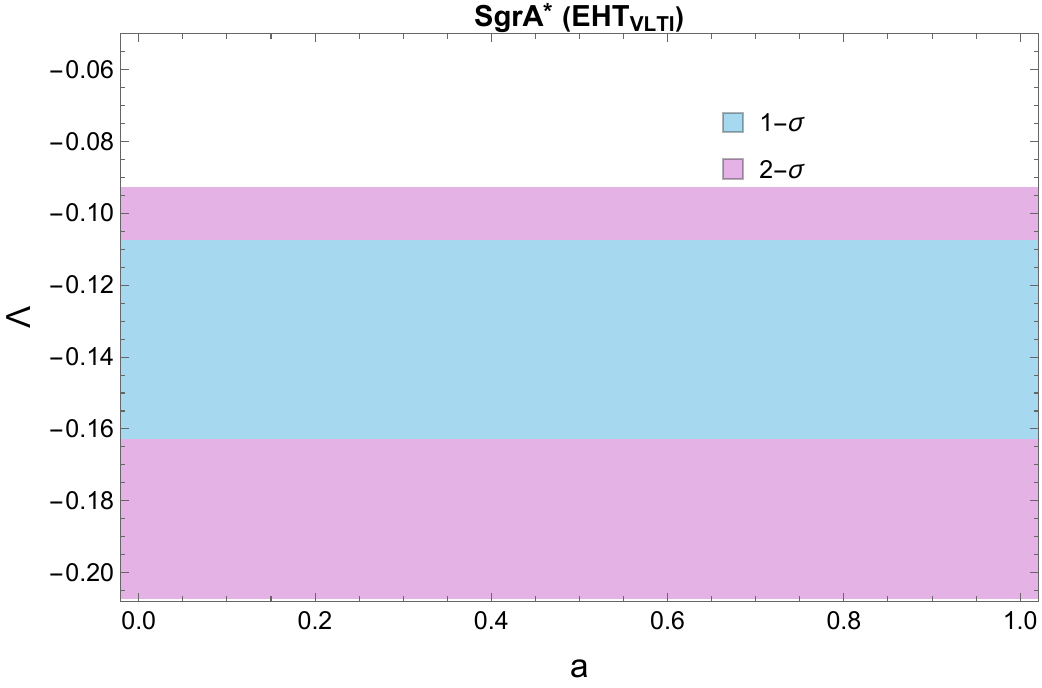}
\includegraphics[scale=0.3]{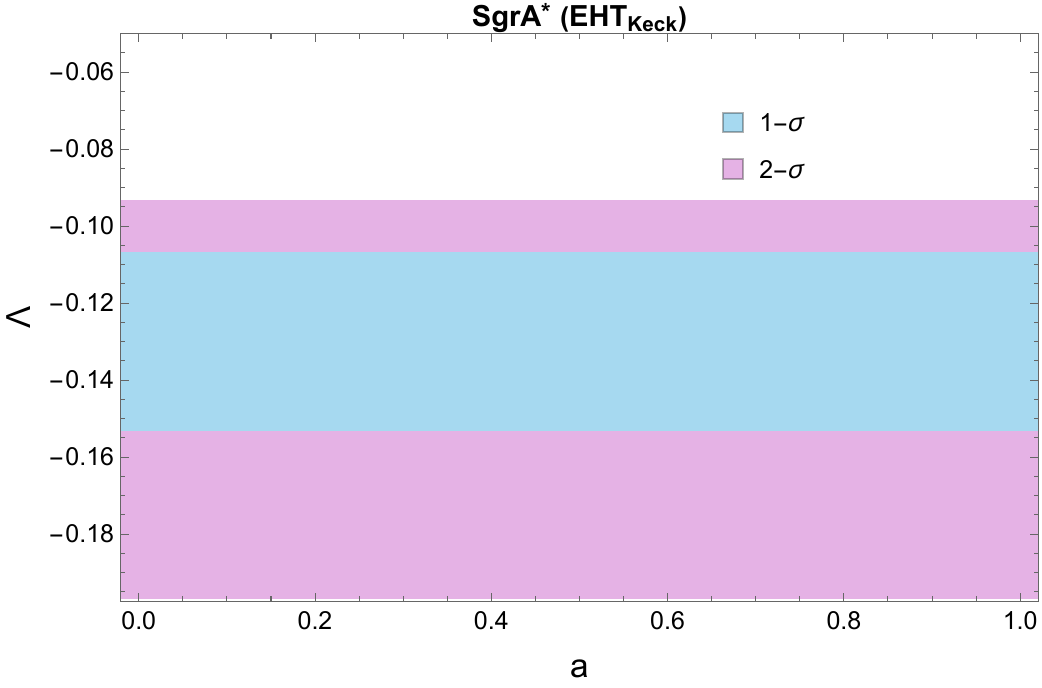}
\caption{\textit{\protect\footnotesize Constraint regions in the $(\Lambda, a)$ plane  by  combining EHT observations and   the Van der Waals limits.}}
\label{ss2}
\end{figure}

\begin{table}[h!]
\centering
\begin{tabular}{|c|c|c|c|}
\hline
\textbf{Black hole} & \textbf{Deviation ($d$)} & \textbf{1-$\sigma$ bounds}
& \textbf{2-$\sigma$ bounds} \\ \hline
M87$^*$ (EHT) & $-0.01^{+0.17}_{-0.17}$ & $4.26 \leq \frac{R_s}{M} \leq 6.03$
& $3.38 \leq \frac{R_s}{M} \leq 6.91$ \\ \hline
Sgr~A$^*$ (EHT$_{\text{VLTI}}$) & $-0.08^{+0.09}_{-0.09}$ & $4.31 \leq \frac{%
R_s}{M} \leq 5.25$ & $3.85 \leq \frac{R_s}{M} \leq 5.72$ \\ \hline
Sgr~A$^*$ (EHT$_{\text{Keck}}$) & $-0.04^{+0.09}_{-0.10}$ & $4.47 \leq \frac{%
R_s}{M} \leq 5.46$ & $3.95 \leq \frac{R_s}{M} \leq 5.92$ \\ \hline
\end{tabular}%
\caption{ \textit{\protect\footnotesize  Fractional deviations and
corresponding bounds for M87$^*$ and Sgr~A$^*$ black holes.}}
\label{t1}
\end{table}
 Now,  we  would like to  explore the  parameter space  regions  producing shadows being compatible with the EHT constraints at the $1-\sigma$ and $2-\sigma$ levels and  the  Van der Waals  limits.   Taking  $M=1$ and varying  the relevant model parameters, we can impose constraints on  the observational bounds on $R_s/M$. The resulting allowed regions in the $(\Lambda,a)$ parameter space regions  are  illustrated in Fig.~\ref{ss2}  at the  Van der Waals limits.
As shown in Fig. (\ref{ss2}), the regions of the reduced parameter space $(\Lambda,a)$ which agree with the empirical observations become larger for increasing parameter values. This result shows that the black hole spacetime can reproduce the observed shadow features. Since $\Lambda$ and $a$ are correlated, we fix one parameter and constrain the other. Taking $a=0.5$, the black hole metric gives a wide range of parameter values with real and physically acceptable horizons. However, the comparison with the EHT observations limits the allowed values of $\Lambda$, and only specific values are consistent with the observed shadow sizes, as follows
\begin{itemize}
\item[•] M87$^*$ case: 
\begin{equation*}
-0.14\le \Lambda \le -0.09 \quad \text{within } 1-\sigma, 
\end{equation*}
\begin{equation*}
-0.27 \le \Lambda \le -0.09, \quad \text{within } 2-\sigma. 
\end{equation*}

\item[•]  Sgr~A$^*$ case  (EHT$_{\text{VLTI}}$): 
\begin{equation*}
-0.165\le \Lambda \le -0.11, \quad \text{within } 1-\sigma, 
\end{equation*}
\begin{equation*}
-0.21 \le \Lambda \le -0.09, \quad \text{within } 2-\sigma. 
\end{equation*}

\item[•]  Sgr~A$^*$  case (EHT$_{\text{Keck}}$): 
\begin{equation*}
-0.15 \le \Lambda \le -0.11, \quad \text{within } 1-\sigma, 
\end{equation*}
\begin{equation*}
-0.17 \le \Lambda \le -0.09, \quad \text{within } 2-\sigma. 
\end{equation*}
\end{itemize}
The results show that, with all other parameters held fixed, $\Lambda$ must be negative and exceed $-0.27$ to remain consistent with the observations.

\section{Machine learning constraints on   black hole parameters by combining EHT  data and Van der Waals behaviors}
Following the analysis of the black hole shadow and its dependence on the model parameters in the Van der Waals limits, it is natural to examine whether the regions of the parameter space compatible with the EHT observations can be identified efficiently without repeatedly performing computationally expensive shadow calculations. This leads  to use machine learning as an alternative approach, capable of capturing the relation between the physical parameters and the observational compatibility of the resulting black hole shadows \cite{s1}. In this framework, we use a fully connected neural network (FCNN) to classify the configurations of the $(\Lambda,a)$ parameter space according to their consistency with the EHT observations.

The datasets are constructed through numerical simulations by exploring the $(\Lambda,a)$ parameter space and comparing the corresponding shadow with the observational constraints of $M87^*$, $Sgr A^*_{\mathrm{VLTI}}$, and $Sgr A^*_{\mathrm{Keck}}$ at both $1$-$\sigma$ and $2$-$\sigma$ confidence levels in the Van der Waals limits. Each configuration is assigned a binary label: Class 1 indicates a shadow consistent with the corresponding observational constraint, whereas Class 0 corresponds to an incompatible configuration. Non-physical and repetitive configurations are excluded. The input parameters are subsequently standardized, and the data are partitioned into training, validation, and test subsets using a fixed random seed, \texttt{random\_state = 42}, thereby ensuring reproducibility.

\begin{figure}[!h]
\centering
\includegraphics[scale=0.25]{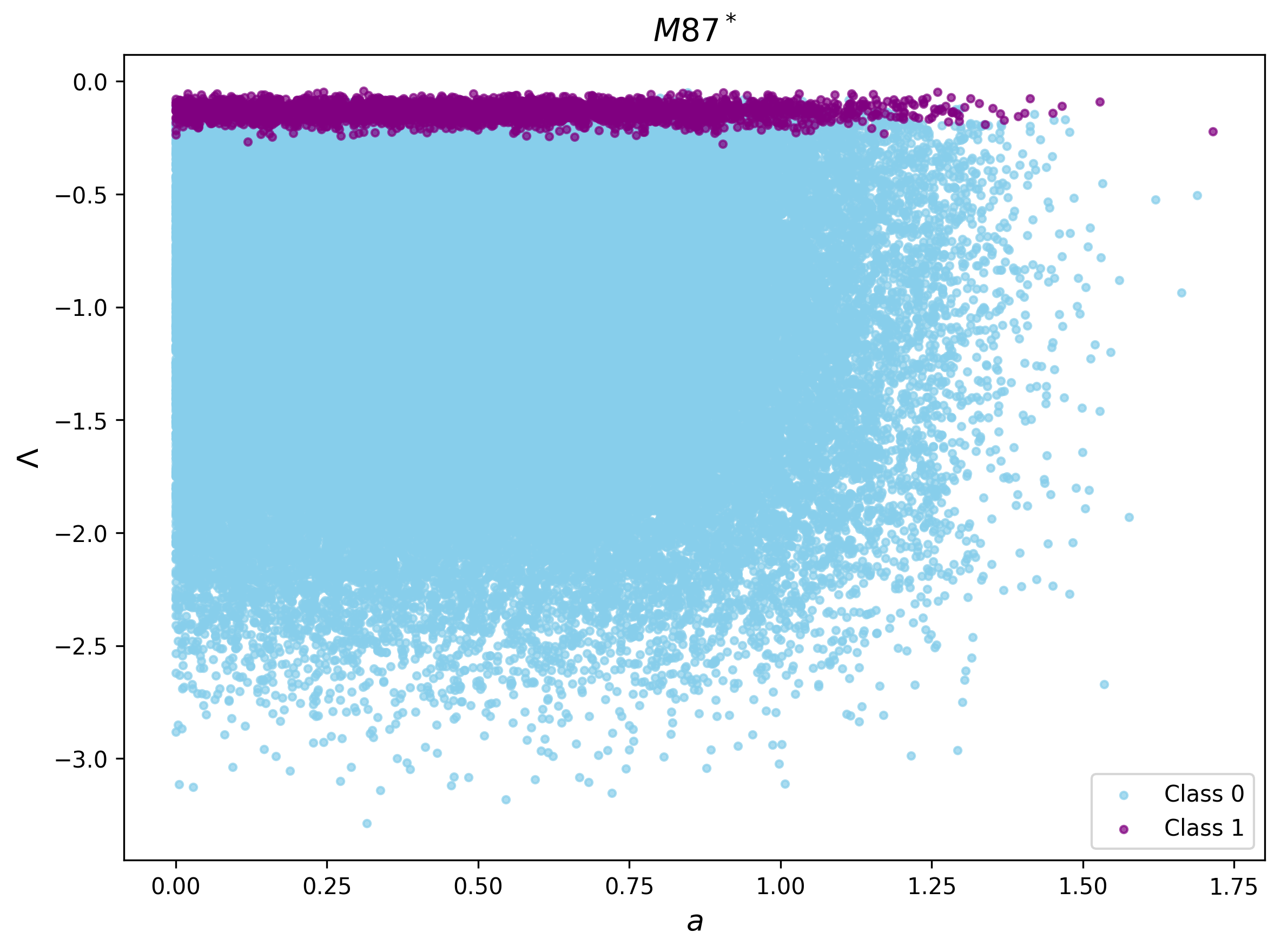}\hspace{2mm}
\includegraphics[scale=0.25]{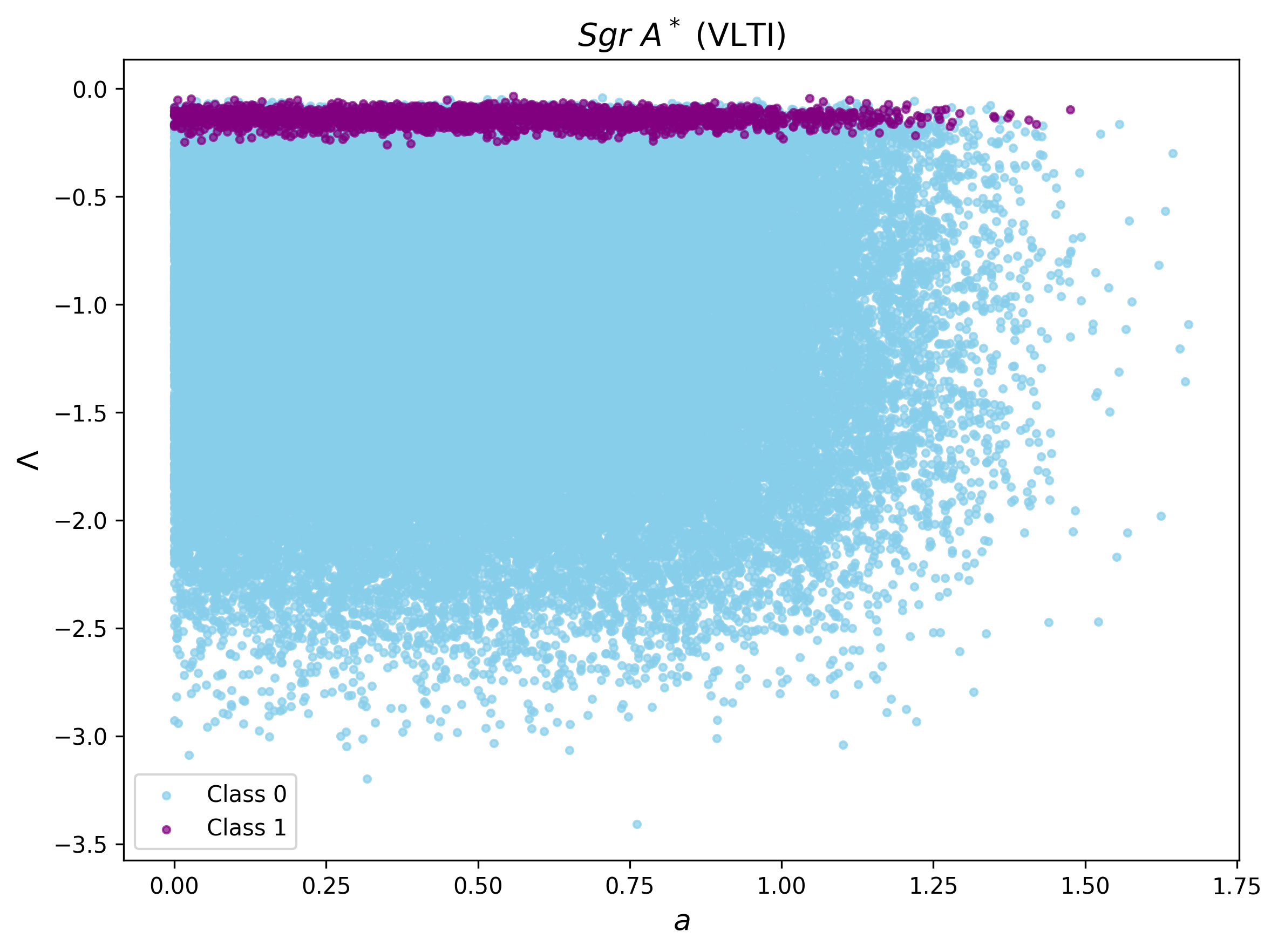}\hspace{2mm}
\includegraphics[scale=0.25]{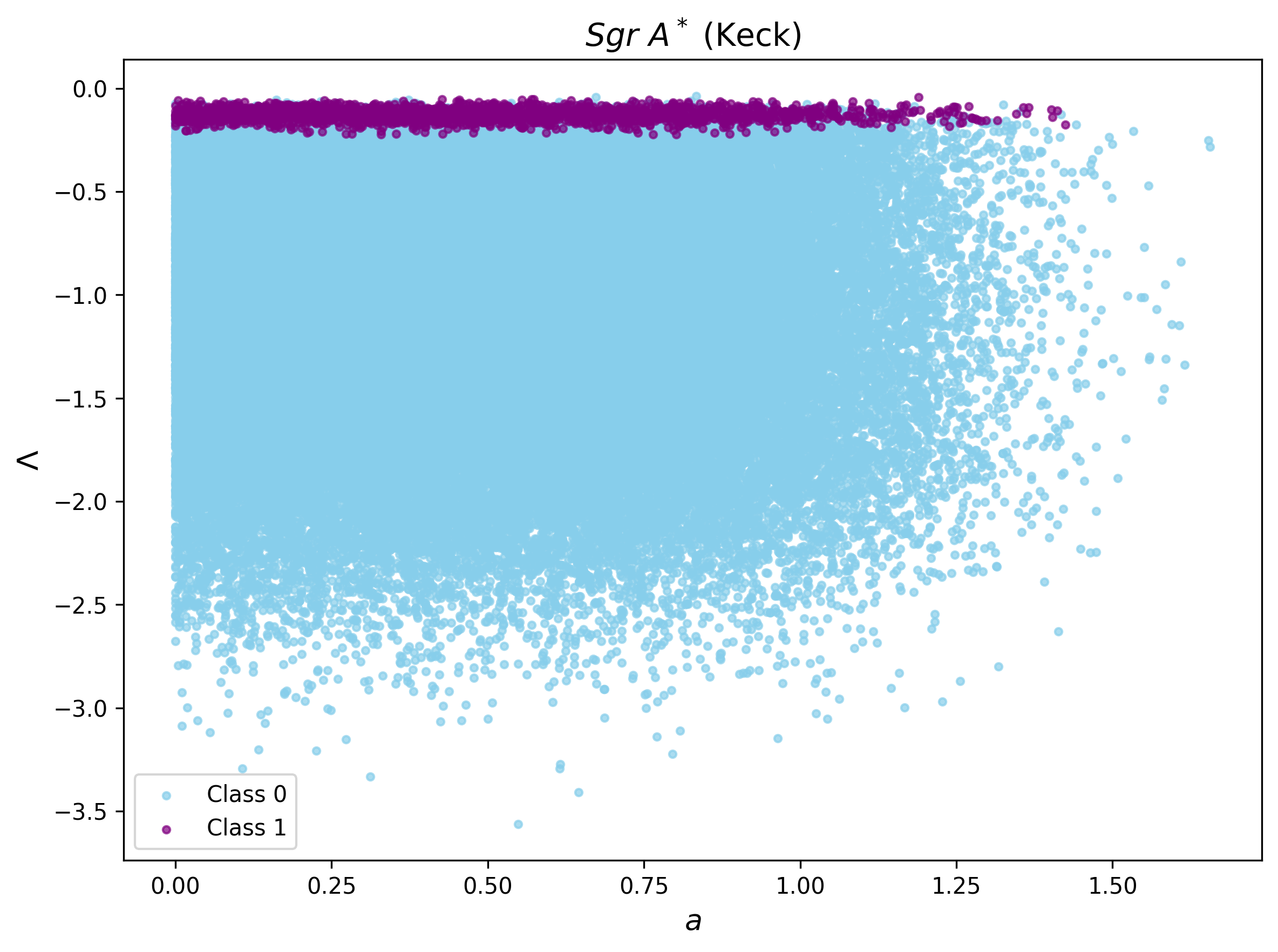}

\includegraphics[scale=0.25]{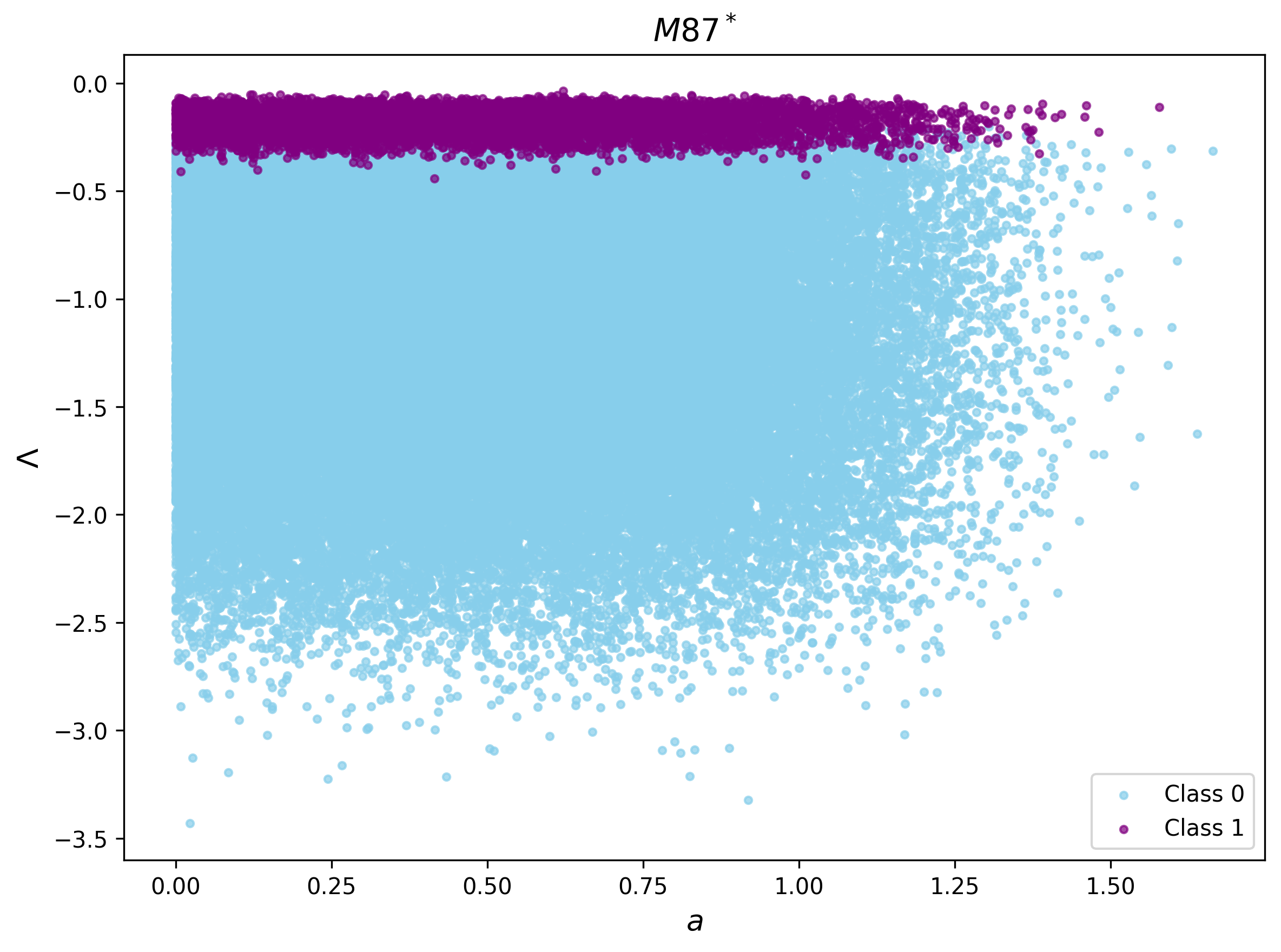}\hspace{2mm}
\includegraphics[scale=0.25]{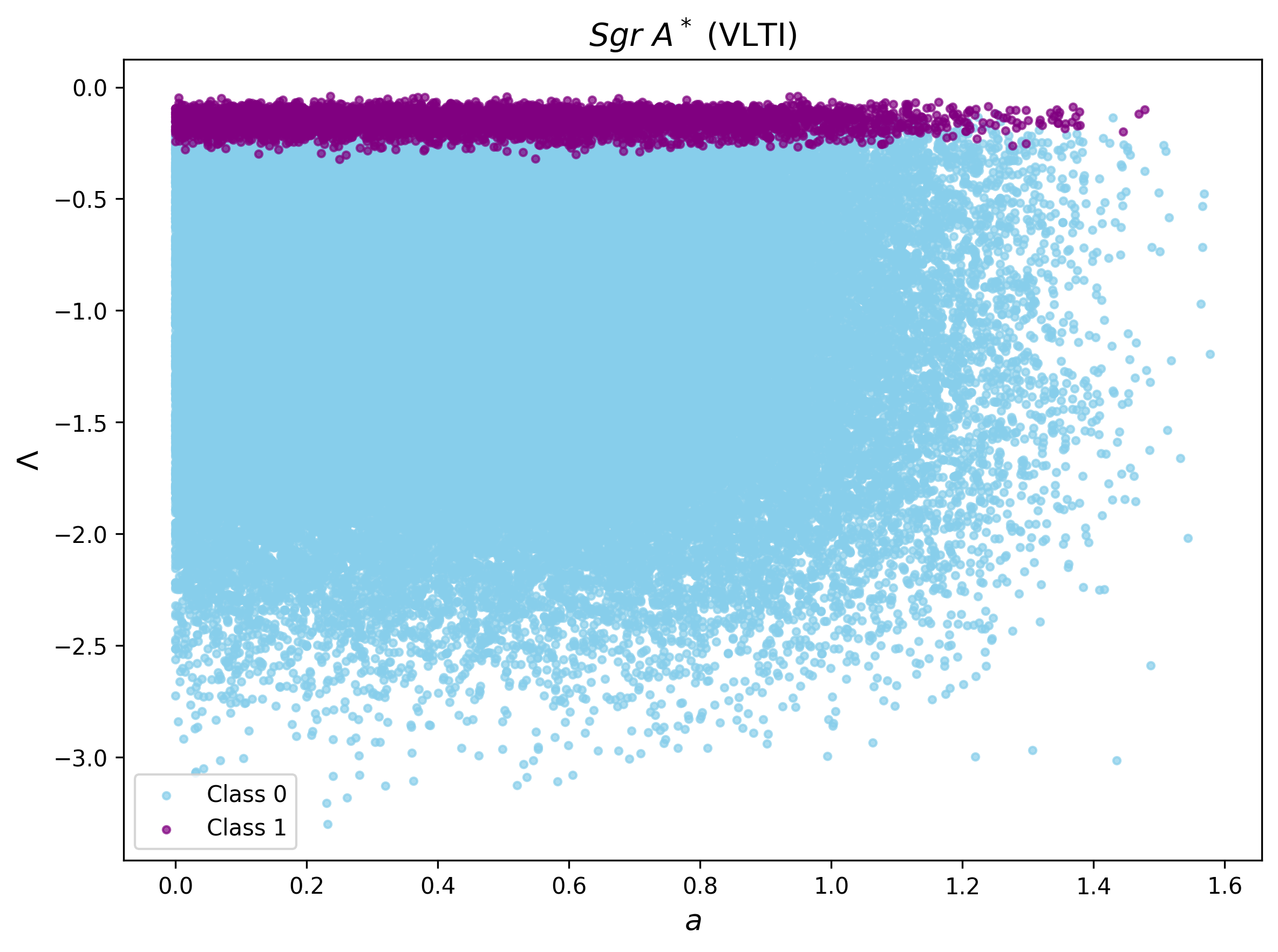}\hspace{2mm}
\includegraphics[scale=0.25]{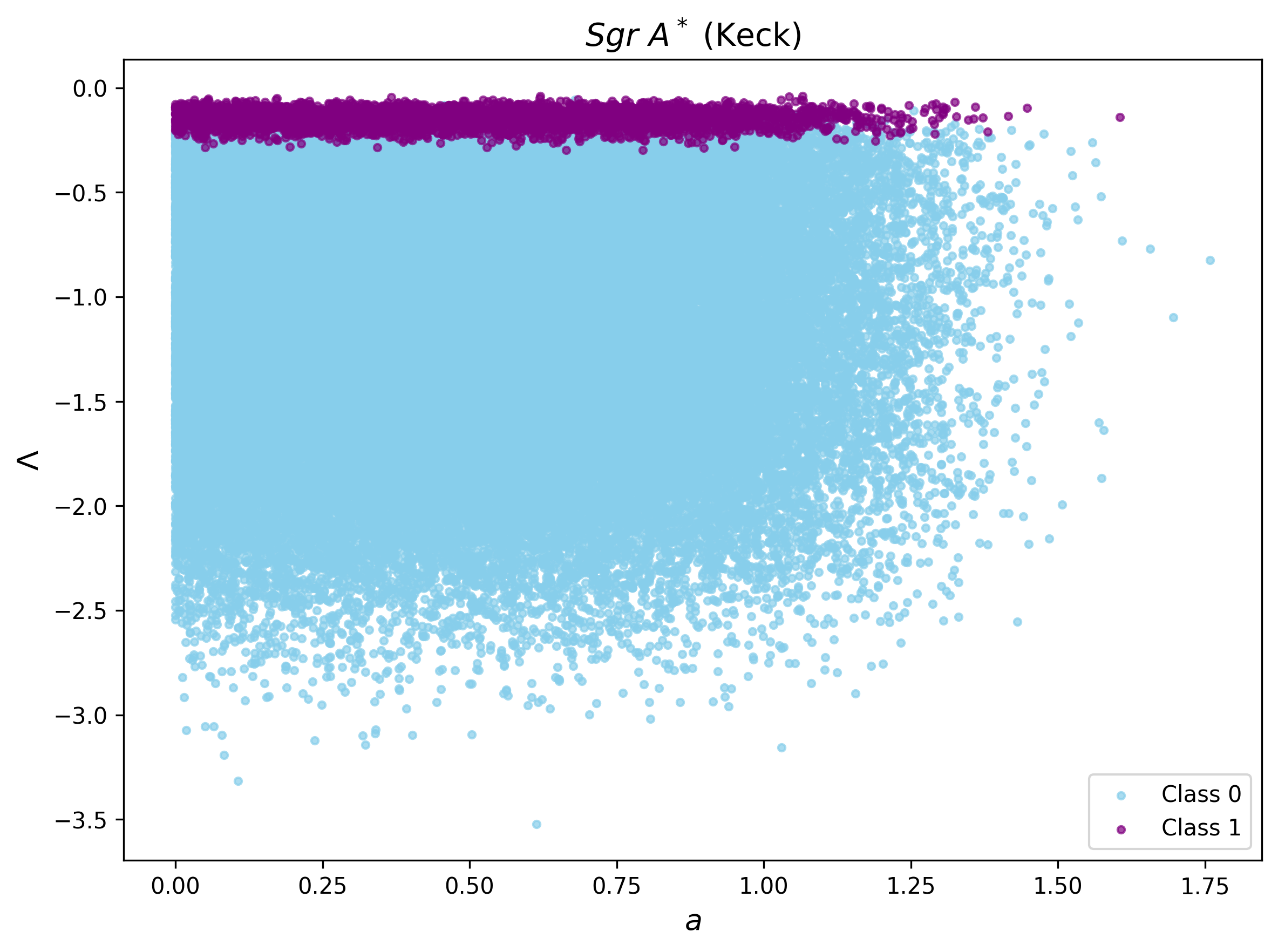}
\caption{Distributions of the generated datasets for $M87^*$, $Sgr A^*_{\mathrm{VLTI}}$, and $Sgr A^*_{\mathrm{Keck}}$ under the $1-\sigma$ (top row) and $2-\sigma$ (bottom row) confidence levels, verifying the Van der Waals behavior.}
\label{fig}
\end{figure}

The resulting distributions are shown in Fig.~(\ref{fig}). Across all observational datasets, the compatible and the incompatible configurations form clearly separated areas within the $(\Lambda,a)$ parameter space, with smooth nonlinear boundaries distinguishing them. The position and the shape of these regions are influenced by both the observational dataset and the confidence level. This organized structure suggests that the shadow compatibility depends on nonlinear relationships among the model parameters, which makes the problem well suited for a supervised learning framework.

The trained FCNN has been evaluated on an independent test set, with the classification accuracy assessed separately for each observational dataset. The results show that the classifier reliably differentiates between the parameter configurations being compatible and incompatible with the EHT observational constraints. This demonstrates that the network successfully detects the associated decision boundary in the reduced moduli space $(\Lambda,a)$. For the black hole models associated with $M87^*$, $SgrA^*_{\mathrm{VLTI}}$, and $SgrA^*_{\mathrm{Keck}}$, the performance indicators corresponding to the $1$-$\sigma$ and $2$-$\sigma$ observational constraints are presented in Table~(\ref{tab1}).

\begin{table}[!ht]
\centering
\caption{Model performance of the FCNN model for the $1-\sigma$ and $2-\sigma$ observational constraints  at the  Van der Waals limits.}
\label{tab1}
\begin{tabular}{l|c|c}
\toprule
 & $1-\sigma$ & $2-\sigma$ \\
 \midrule
{Metric} 
& $M87^*$ \quad $SgrA^*_{\mathrm{VLTI}}$ \quad $SgrA^*_{\mathrm{Keck}}$
& $M87^*$ \quad $SgrA^*_{\mathrm{VLTI}}$ \quad $SgrA^*_{\mathrm{Keck}}$ \\
\midrule
Test Accuracy 
& 0.9844 \quad 0.9784 \quad  0.9833
& 0.9736 \quad 0.9818 \quad  0.9816\\
\bottomrule
\end{tabular}
\end{table}

To assess the robustness of the trained classifier, we employ a Monte Carlo voting strategy based on Gaussian perturbations of the normalized input features. By adding Gaussian noises with a standard deviation of $0.01$ to the normalized input vector, we generate 20 perturbed realization for each sample in the independent test set. The perturbed realization is then independently evaluated by the FCNN, and the final prediction is obtained through majority voting among the 20 outputs. These perturbations are included only after the training phase, therefore they do not affect the target labels produced by the numerical simulations. This procedure provides a measure of the classifier  stability in the neighborhood of each test sample and reduces its sensitivity to small numerical perturbations.

Considering the $M87^*$, $SgrA^*_{\mathrm{VLTI}}$, and $SgrA^*_{\mathrm{Keck}}$  black hole models, the confusion matrices corresponding to the $1$--$\sigma$ and $2$--$\sigma$ observational constraints at the Van der Waals limits are  shown  in Table~(\ref{c11}). The results reveal  that the FCNN correctly classifies the vast majority of samples, with only a very small number of false positives and false negatives.

\begin{table}[!ht]
\centering
\caption{Confusion matrix for FCNN model  with  $1-\sigma$ and $2-\sigma$ observational constraints  at the  Van der Waals limits.}
\label{c11}
\begin{tabular}{l|l|c|c|c}
\toprule
\textbf{} & \textbf{Class} & \textbf{$M87^*$} 
& \textbf{$SgrA^*_{\mathrm{VLTI}}$} & \textbf{$SgrA^*_{\mathrm{Keck}}$} \\
\midrule
\multirow{2}{*}{$1-\sigma$} 
&  & Class 0 \quad Class 1 &  Class 0 \quad Class 1 &  Class 0 \quad Class 1 \\
& Class 0 &14,682 \qquad  146&14,775\qquad 224 &14,755 \qquad 193  \\
& Class 1 &94 \qquad 471& 43 \qquad 361 &139 \qquad 306  \\
\midrule
\multirow{2}{*}{$2-\sigma$} 
&  & Class 0 \quad Class 1 &  Class 0 \quad Class 1 &  Class 0 \quad Class 1 \\
& Class 0 &13,897 \qquad 163&14,443\qquad 183 &14,455 \qquad 89  \\
& Class 1 &242 \qquad 1,019 &99 \qquad 668&191 \qquad 658  \\
\bottomrule
\end{tabular}
\end{table}
A close examination reveals that the classifier performs considerably better under $1-\sigma$ observation constraints than under $2-\sigma$ constraints.  It has been observed that the $M87^*$ dataset consistently reaches the highest classification accuracy. This difference can be interpreted as the $1 -\sigma$ confidence interval providing a more restrictive compatibility range. This results in a more visible separation between the two classes.  Nevertheless,   the $2 -\sigma$ confidence interval  enlarges  the region of acceptable parameter configurations. This makes the classification   more difficult.  However, the model maintains a highly accurate performance across all datasets, revealing its strength and capacity to effectively generalize to different empirical confidence levels from the EHT observations.

\section{Conclusion}
In this work, we have  investigated certain physical behaviors of  NC  Schwarzschild  black hole solutions with GMs from  Van der Waals-like  features. In particular, we  have recovered such behaviors by approaching the $(P-v) $ criticality and   the  Joule--Thomson effect. Specifically,  such  a thermodynamic  limit  has been combined  with  the  optical computations of the black holes    to  examine the associated shadows.  Employing  such calculations, we have illustrated shadows properties in terms of circular curves in  two dimensions.   Using   EHT observational data, we  have constrained the  associated  black hole parameters including the NC one. After that, by combining machine learning methods with empirical data for the black holes $M87^*$, $Sgr A^*_{\mathrm{VLTI}}$, and $Sgr A^*_{\mathrm{Keck}}$, and evaluating the predictions within the $1-\sigma$ and $2-\sigma$ confidence intervals reported by the EHT  international collaborations, we have established  black hole parameter constraining scenarios.  Precisely,  we have found that the proposed black hole  model in NC geometry with GMs   matches with   the $M87^*$ observational  data. 

This work leaves certain open questions.  A natural question is to go beyond the neutral and  non-rotating  solutions.  In fact, an extension of the present work could be addressed in future works by implementing extra internal and external parameters.

{\bf Data availability}\\
  The data are available from the corresponding author upon reasonable request. \\

\section*{Acknowledgements}   
MJ would like to thank A. Belhaj for his discussions, suggestions, guiding,  and  supervising.  She  gratefully acknowledges the financial support of the CNRST in the frame 
of the PhD Associate Scholarship Program PASS.

\end{document}